\documentclass[journal]{IEEEtran}
\ifCLASSINFOpdf
\else
\fi
\usepackage{graphicx}
\usepackage[utf8]{inputenc}
\usepackage{siunitx}
\usepackage{booktabs} 
\usepackage{array}
\begin{document}
%
\title{Spectroscopic Signature of Local Alloy Fluctuations in InGaN/GaN Multi-Quantum-Disk Light Emitting Diode Heterostructures and Its Impact on the Optical Performance}
%
%
%

\author{Soumyadip~Chatterjee,
        Subhranshu~Sekhar~Sahu,
        Kanchan~Singh~Rana, 
        Swagata~Bhunia,
        Dipankar~Saha,
        and~Apurba~Laha,
\thanks{S. Chatterjee, K. S. Rana, D. Saha, and A. Laha are with the Department
of Electrical Engineering, Indian Institute of Technology Bombay,
Mumbai, 400076 India (e-mail: laha@ee.iitb.ac.in).}
\thanks{S. S. Sahu and S. Bhunia are with the Department
of Physics, Indian Institute of Technology Bombay,
Mumbai, 400076 India}
\thanks{Manuscript received XXXX XX, XXXX; revised August XX, XXXX.}}

%
%

\markboth{XXXX XXXX of XXXX,~Vol.~XX, No.~X, XXXXX~XXXX}%
{Shell \MakeLowercase{\textit{et al.}}: Bare Demo of IEEEtran.cls for IEEE Journals}
%



\maketitle

\begin{abstract}
Inhomogeneity-governed carrier localization has been investigated in three sets of InGaN/GaN multi-quantum-disk light-emitting diode (LED) structures grown by plasma-assisted molecular beam epitaxy (PAMBE) under different process conditions. A temperature-dependent study of the luminescence peak positions reveals that samples prepared under certain process conditions exhibit a thermal distribution of carriers from the localized states that show the typical 'S' shaped dependence in luminescence characteristics. The absence of an 'S' shaped nature in the other sample prepared with relatively higher In-flux infers a superior homogeneity in alloy composition. Further investigation manifested superior optical properties for the samples where the 'S' shape nature is found to be absent.
\end{abstract}


%
\IEEEpeerreviewmaketitle

\section{Introduction}
%
%
%
%
\IEEEPARstart{E}{pitaxially} grown InGaN/GaN nanowire light-emitting diodes (LED) have disrupted the field of optoelectronics in recent years owing to their structurally pure nature and bandgap tunability, among other advantages \cite{ZHAO201514}. Especially the axial nanowire LED structures, since its early days \cite{Kikuchi_2004}, have seen enormous development, potentially emerging as a suitable technology for realizing full-color displays \cite{NW_display}. Thus, several structural, electrical, and optical properties of these structures have been studied thoroughly in recent years \cite{disc-in-wire_properties, D2NR05529E, nanoscale_led, MNHASAN2022126654, Kehagias_2013, MQW_NW_LEDs}. 

One critical aspect of the InGaN/GaN heterostructures is the presence of localized states due to random compositional fluctuations \cite{localization_nakamura, stong_local_apl}. Mostly strain, surface mobility aside few other factors, causes these fluctuations \cite{DUBOZ2023127033}. The presence of these localized states can be observed through the signature 'S'-shaped temperature dependence of luminescence peaks \cite{Badcock_2013, S_APL, S_anm}. These localized states help enhance the luminescence efficiency of the device \cite{GRANDJEAN2000495,enhance_efficiency}. However, studies also suggest too much inhomogeneity can cause inferior optical performance in high In-concentration samples, especially for emissions in green wavelength regions and higher \cite{PhysRevLett.116.027401}. Incidentally, almost all the work concerning localized state and the S-shaped temperature dependence of peak positions are confined to the planar InGaN layers \cite{thermal_redistribution} and InGaN/GaN quantum wells \cite{Li_2015, LIU2015266}. Such studies for InGaN/GaN axial nanostructures are exceptionally rare. Although some reports show the peak energy in quantum disk structures follows Varshini's law \cite{LED_on_metal, red_mqd_LED} instead of the S-shaped nature, no comprehensive study on this topic exists to date. No work, to the best of our knowledge, reports correlating localization-dependent peak shift with growth conditions, analyzing its implication on the optical performance of the nanostructure. However, such a study is immensely important for understanding the role of compositional fluctuations in the InGaN disk-in-wire structure, especially for higher Indium concentrations ($>$20\%).

In this work, we have grown three sets of InGaN/GaN multi-quantum disk LED structures using the PAMBE system under three different process conditions. We have carried out temperature-dependent photoluminescence (PL) studies to study the impact of localization states in each sample through thermal redistribution of excitonic carriers. Finally, power-dependent PL and time-correlated single photon count (TCSPC) studies have been carried out to investigate further the impact of alloy fluctuation-related localization on the optical properties.

\section{Experimental Section}
The samples used in this study were grown on Si(111) substrates in a Riber C21 MBE system equipped with a Veeco Unibulb plasma cell. Before the growth, RCA-cleaned Si wafers were annealed in the growth chamber at 850{\textdegree}C for 1 hr till surface reconstruction. The surface of the wafer was then nitridated at 800{\textdegree}C for 15 mins to enable steady nucleation of GaN nanowires on the substrate \cite{EFTYCHIS20168}. Si-doped GaN nanowires were grown for 90 mins on these nitridated substrates at 780{\textdegree}C, with a Ga-flux of $2.6\times10^{-7}$ Torr and 430 W N$_2$ plasma with a flow rate of 2.3 sccm. On top of these nanowires, 7 pairs of InGaN/GaN well-barrier structures were grown. The growth conditions of these heterostructures were chosen appropriately to enable luminescence in the 'green gap' \cite{PhysRevLett.116.027401} region. The growth temperature and the fluxes are shown in Table \ref{tbl:1}. 
\begin{table*}[ht!]
\caption{\label{tbl:1}Growth condition for InGaN/GaN multi quantum-disk regions}
\footnotesize
\centering
\begin{tabular}{@{}lllll}
\toprule
Sample Name&Growth Temperature&In-flux (Torr)&Ga-flux (Torr)&Nitrogen Plasma\\
\midrule
Sample A&550 {\textdegree}C&$7.3\times10^{-8}$&$1.2\times10^{-7}$&430 W, 2.3 sccm\\
Sample B&525 {\textdegree}C&$7.3\times10^{-8}$&$1.2\times10^{-7}$&430 W, 2.3 sccm\\
Sample C&550 {\textdegree}C&$1.4\times10^{-7}$&$1.2\times10^{-7}$&430 W, 2.3 sccm\\
\bottomrule
\end{tabular}
\end{table*}
\normalsize

On top of these multi-quantum-disk structures, an Mg-doped AlGaN electron-blocking layer was grown, followed by the growth of an Mg-doped GaN layer at 650{\textdegree}C. A field emission gun-scanning electron microscope (FEG-SEM) and a 300 keV transmission electron microscope (TEM) were employed for the structural characterization of the samples. The PL measurements were done in a temperature-dependent PL setup with a He-Cd Laser (325 nm). The transient luminescence measurements were done in a TCSPC setup with a femtosecond pulsed 325 nm LASER source.

\section{Results and Discussion}
\begin{figure}[b]
  \includegraphics[width=0.48\textwidth]{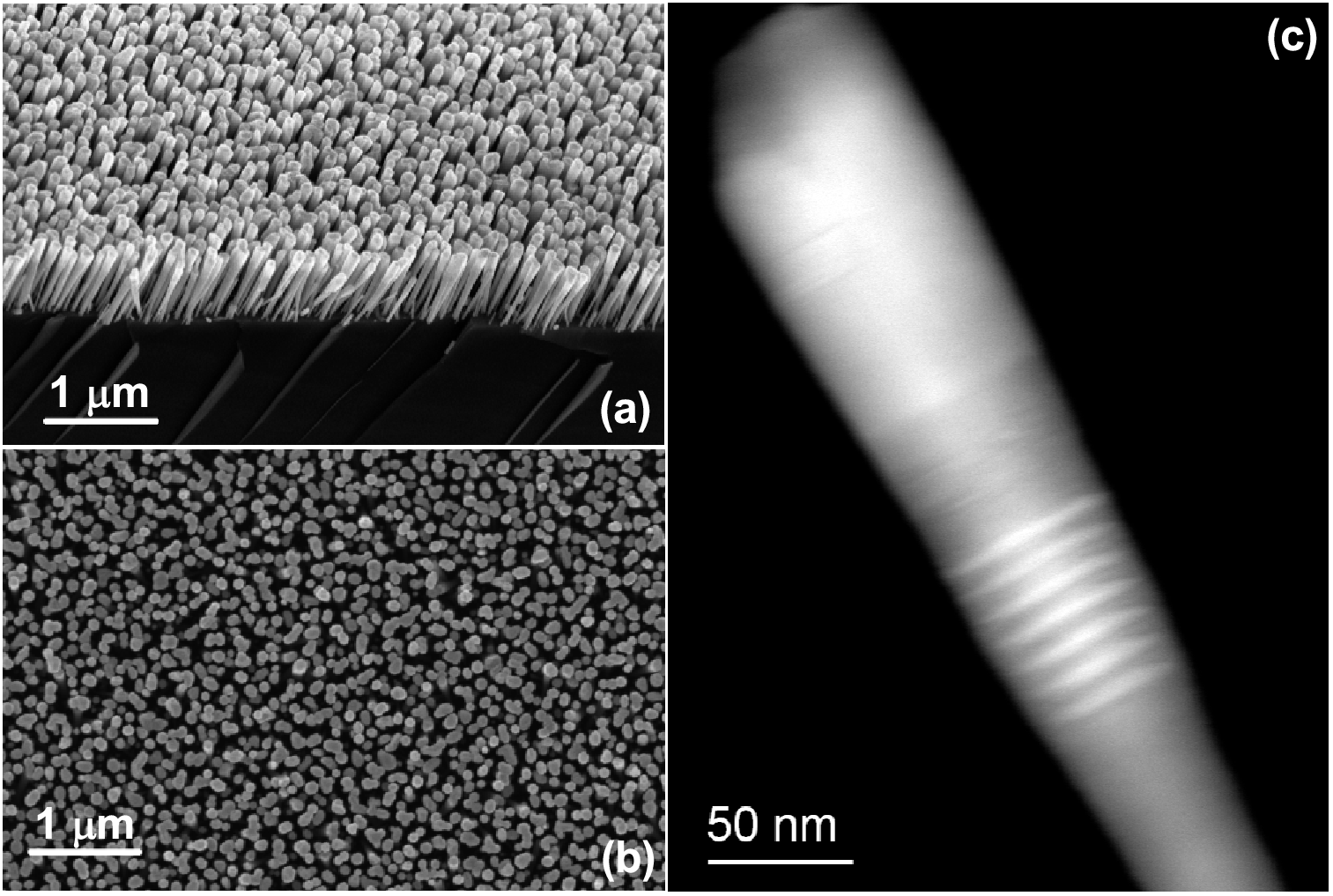}
  \caption{(a) 45{\textdegree} tilted view and (b) top view FEG-SEM image of an InGaN/GaN multi quantum-disk LED sample grown on Si(111). (b) Annular dark-field STEM image of an isolated nanowire with the 7 bright disk-like regions being the 7 InGaN disks with GaN barriers (darker region) between them.}
  \label{fgr:1}
\end{figure}

FEG-SEM images confirmed the realization of self-assembled nanowires. Figure \ref{fgr:1}(a) and \ref{fgr:1}(b) show the 45{\textdegree} tilted and top view FEG-SEM images of a nanowire LED sample. The length of the nanowires is around 800 nm. An annular dark-field scanning transmission electron microscopy (STEM) image of an individual nanowire verified the formation of the seven quantum disks, as shown in figure \ref{fgr:1}(c). 

\begin{figure}[t]
  \includegraphics[width=0.48\textwidth]{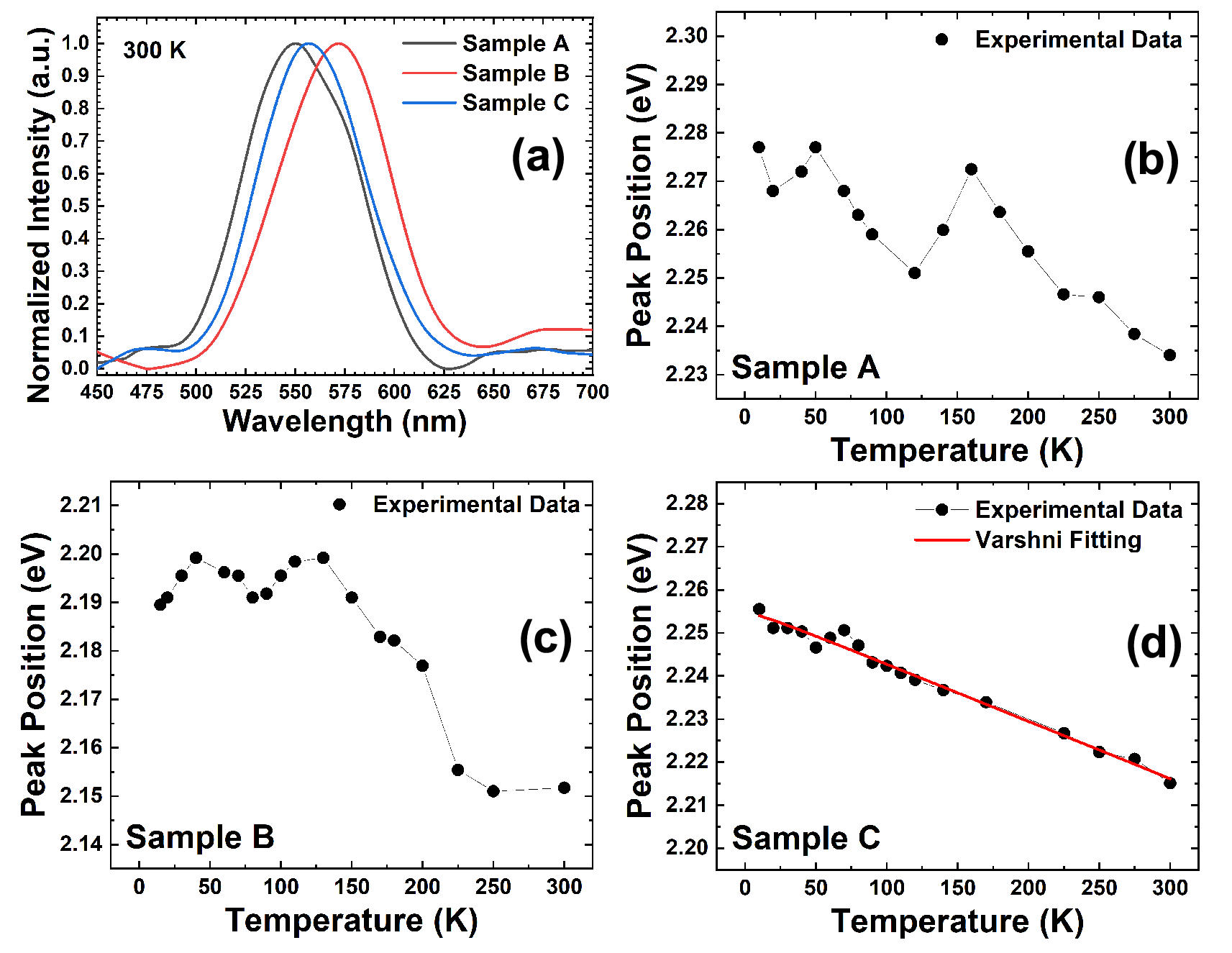}
  \caption{(a) Photoluminescence spectra of the samples at 300K, normalized with respect to characteristic peaks. Characteristic peak emission energy from PL spectra as a function of temperature and for (b) Sample A, (c) Sample B, and (d) Sample C}
  \label{fgr:2}
\end{figure}

The optical characterizations of the samples were done through temperature-dependent PL measurement. From the PL spectra captured at 300 K, we observed InGaN-related luminescence peaks for all the samples to be inside the green gap region, as can be seen in the normalized spectra in figure \ref{fgr:2}(a). The characteristic peaks are at 554 nm, 576 nm, and 560 nm for sample A, sample B, and sample C, respectively. The higher wavelength in sample B can be attributed to the lower growth temperature of the quantum-disk regions, aiding higher In-incorporation by reducing the decomposition rate of InGaN. Similarly, a higher In-flux in sample C aids slightly higher In-incorporation than in sample A. These results highlight the dominance of the decomposition and desorption processes during the In-incorporation in an InGaN/GaN nanostructure. 

\begin{figure}[t]
  \centering
  \includegraphics[width=0.48\textwidth]{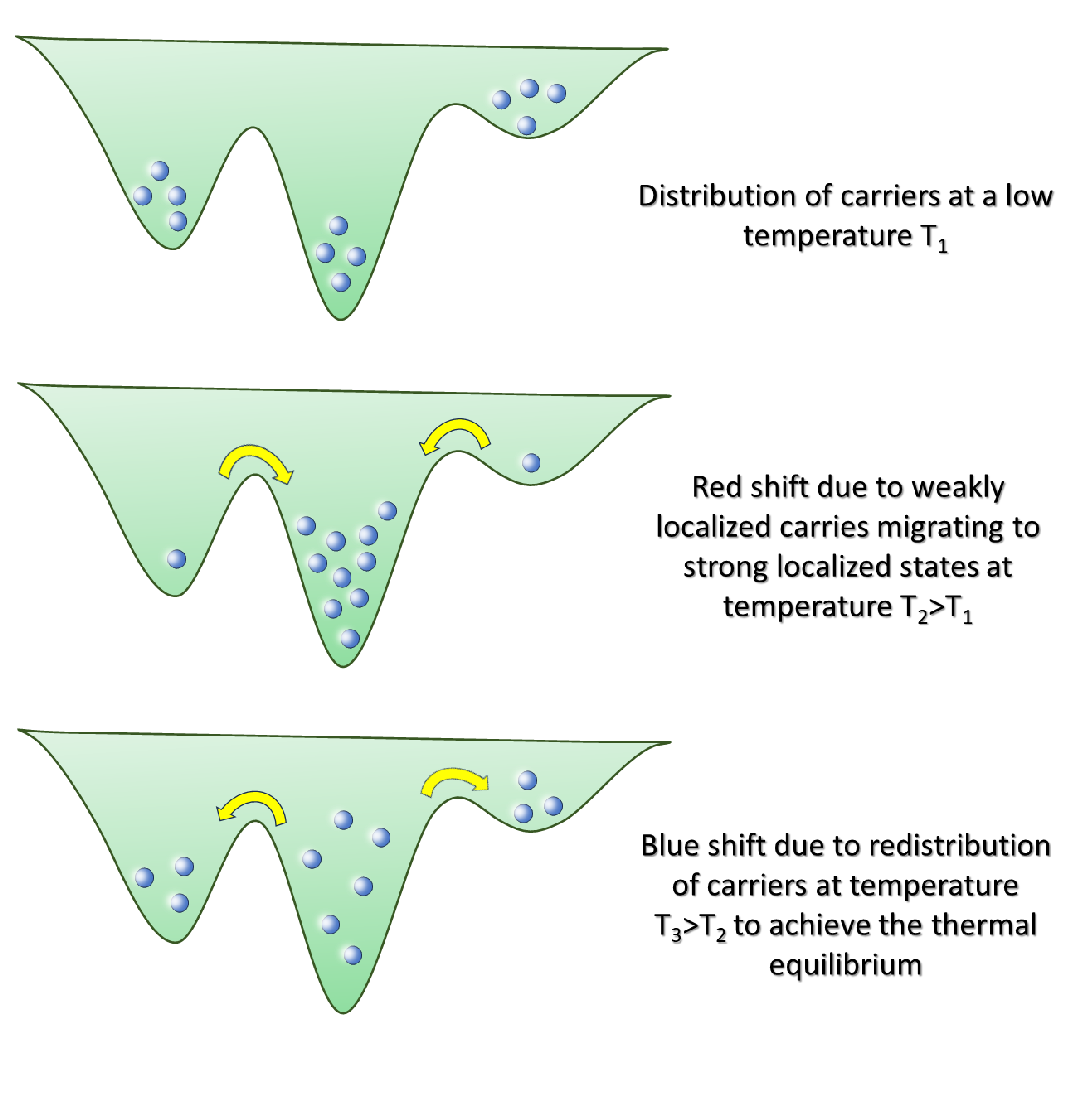}
  \caption{Mechanism of thermal redistribution of carriers in the vicinity of a localization center explaining characteristic red shift and blue shift of emission peak with temperature}
  \label{fgr:3}
\end{figure}

The peak positions of temperature-dependent PL spectra were estimated in terms of emission energy to investigate the localization and delocalization of carriers within the nanostructures. The estimated peak positions as a function of temperatures are shown in figure \ref{fgr:2}(b), figure \ref{fgr:2}(c), and figure \ref{fgr:2}(d). It can be clearly seen from the results that the peak positions for sample A and sample B do not follow the usual nature of semiconductor bandgaps predicted by Varshni's equation \cite{VARSHNI1967149} as below
\begin{equation}
E_g(T)=E_g(0)-\frac{{\alpha}{T^2}}{\beta+T}\label{eqn:1}
\end{equation}
Here, T is the temperature in the Kelvin scale, and $E_g(0)$ is the band gap at T=0 K. The constants $\alpha$ and $\beta$ are Varshni's thermal coefficients. Only sample C shows a signature red shift of emission peaks with temperature, following Varshni's law. The phenomenon behind the anomalous temperature dependence of peak position in samples A and B is elucidated using figure \ref{fgr:3}.

At lower temperatures, photogenerated carriers are localized at their nearest localization state as they don't get sufficient thermal energy to overcome the potential barriers. As the temperature increases, the carriers gain enough energy to overcome the barrier and relax in the strongly localized states. This results in an initial redshift of the peak positions. As the temperatures increase further, carriers tend to move to higher energy bands, eventually redistributing them among the localization centers. This manifests in the anomalous blue shift of peak position. Once the carriers are redistributed, the peak position starts following the characteristic redshift as predicted in Varshini's law.

All these phenomena are inherently dependent on the alloy fluctuations in the samples, as the localization states originate from the Indium-rich clusters acting similar to quantum dots \cite{Badcock_2013, S_APL, S_anm}. Therefore, a stronger 'S'-shaped nature results from higher inhomogeneity in the InGaN layers. In the case of sample A, the carrier redistribution to higher energy levels (higher energy shift of the peak position) starts at around 120 K, unlike sample B, where it starts around 80 K. This suggests the presence of deeper localization states (higher In-concentration difference between the layer and clusters) in sample A in comparison to the other samples. Similarly, the temperature-dependent peak position plot of sample C has barely any signature of any localization/delocalization process, suggesting a relatively homogenous composition of In in the InGaN disk regions. 

\begin{figure*}[t]
  \includegraphics[width=\textwidth]{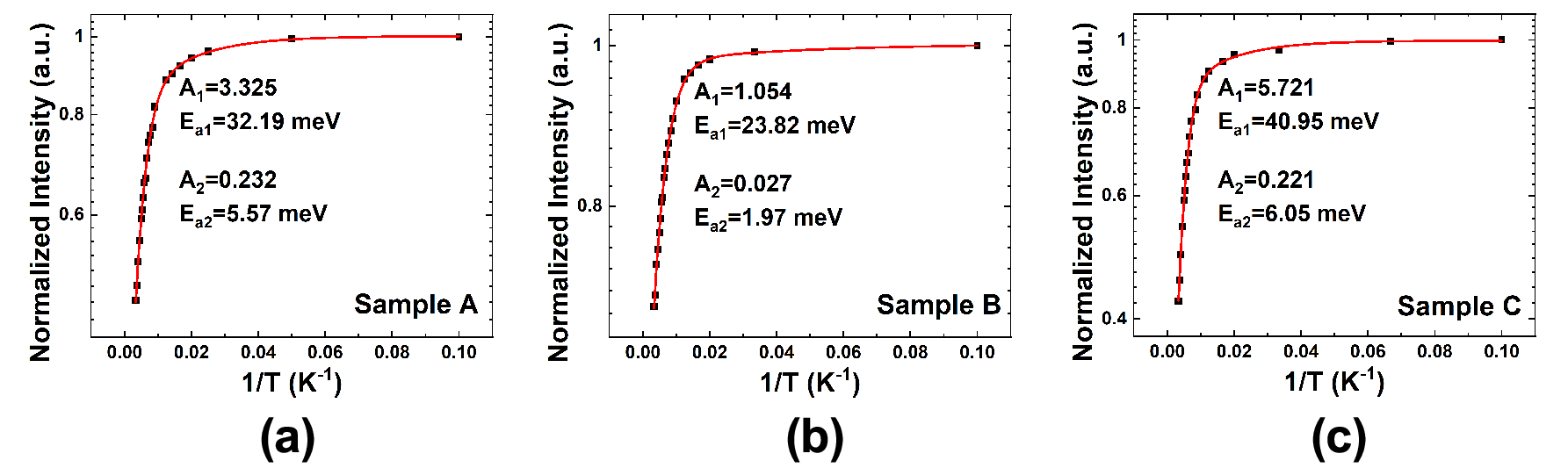}
  \caption{Integrated PL intensity as a function of temperature for (a) sample A, (b) sample B, (c) sample C and their fitting based on Arrhenius equation}
  \label{fgr:4}
\end{figure*}

The implications of this inhomogeneity in the these samples were explored by studying the defect-related thermal quenching properties \cite{th_quench}. The intensity of PL peaks is higher at low temperatures, and they reduce gradually as more and more carriers recombine through the non-radiative recombination, causing thermal quenching. Each non-radiative path has an activation energy associated with it, which is estimated using the Arrhenius equation \cite{Arrhenius_InGaN} shown below
\begin{equation}
I(T)=\frac{I_0}{1+{\sum_{i}}{A_i}exp(-\frac{E_{ai}}{k_BT})}
\label{eqn:2}
\end{equation}
I(T) represents the PL intensity at a temperature T, $I_0$ is the PL intensity at a low temperature, and $k_B$ is the Boltzmann constant. $A_i$ and $E_{ai}$ represent the rate constants and activation energies of an individual non-radiative recombination channel i. Figure \ref{fgr:4} shows the plot of the normalized PL intensities for the samples and the activation energies estimated from the Arrhenius equation. 

Interestingly, the best fitting is achieved considering the presence of two separate non-radiative recombination channels in the structure. Studies mostly associated the channel with lower activation energy (typically $<$10 meV) with interface-related defects \cite{Tian_2017, Interface_activation} and the one with higher activation energy with dislocation-related defects \cite{Tian_2017, Dislocation_PL}. It can be observed from the results that sample C has the highest values of activation energy (Figure \ref{fgr:4}(c)), manifesting an improved condition for radiative recombination. Sample B, grown at the lowest growth temperature (525 {\textdegree}C), shows the lowest activation energies (Figure \ref{fgr:4}(b)). Usually, the edge and screw dislocation are more prominent in the vicinity of an alloy disorder in a crystal. This happens as the segregation of Indium adatoms during the growth of InGaN layers is more energetically favorable around dislocation cores \cite{segre_dislo}. The dislocation sites also allow the Indium-rich clusters to relieve tensile strain. This is why sample C, with a relatively homogeneous structure, also possesses an enhanced activation energy $E_{a1}$ (a signature of superior crystalline quality). Since the In adatoms enhance the mobility of Ga adatoms \cite{Taib_2023, In_surfactant}, the higher Indium flux used during the growth of sample C can be considered a pivotal factor for this improved crystalline quality. 

On the other hand, since both barrier and well are grown at the same temperature, the interface quality is expected to be better for higher growth temperatures. This is why samples A and C have similar values of $E_{a2}$, while for sample B, $E_{a2}$ is much lower at 1.97 meV.

\begin{figure*}[t]
  \includegraphics[width=\textwidth]{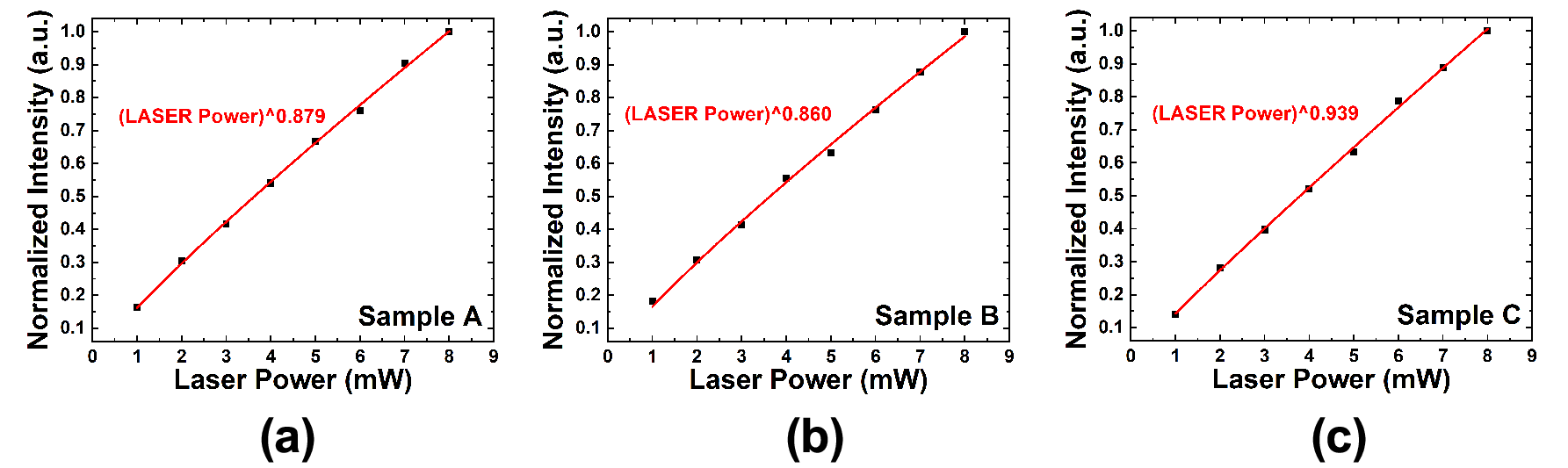}
  \caption{Variation of PL intensity with the LASER power at 10 K for (a) sample A, (b) sample B, and (c) sample C with lower exponents of LASER power suggesting higher non-radiative recombination rate}
  \label{fgr:5}
\end{figure*}

The lower defect density in sample C is also evident from the power-dependent PL intensity study of the samples at 10 K, as shown in figure \ref{fgr:5}. If we consider I as the intensity of PL peaks and P the excitation LASER power, we can relate the quantities through exponent f using the relation \cite{Nag_2020}
\begin{equation}
I \propto P^f
\label{eqn:3}
\end{equation}

In the absence of nonradiative paths, all the photo-injected carriers contribute to the luminescence intensity, resulting in the linearity of the relation in eqn \ref{eqn:3} with the value of f reducing to unity. However, when defects are present, some of the injected carriers are lost through non-radiative recombination, especially at higher carrier concentrations at high excitation powers. This results in a smaller value of the exponent f ($<$1). 

It is well established that despite localization, some carriers can still tunnel to non-radiative centers even at cryogenic temperatures \cite{SRH_tunnel, Yu_2022}. Thus, a higher density of defects results in a smaller value of f. In this case, sample C also shows the highest value of f at 0.939 (figure \ref{fgr:5}(c)), further validating the notion of lower defect density in that sample. This provides additional evidence regarding the connection between alloy inhomogeneity and dislocation defects. 

  \begin{figure}[t]
  \includegraphics[width=0.48\textwidth]{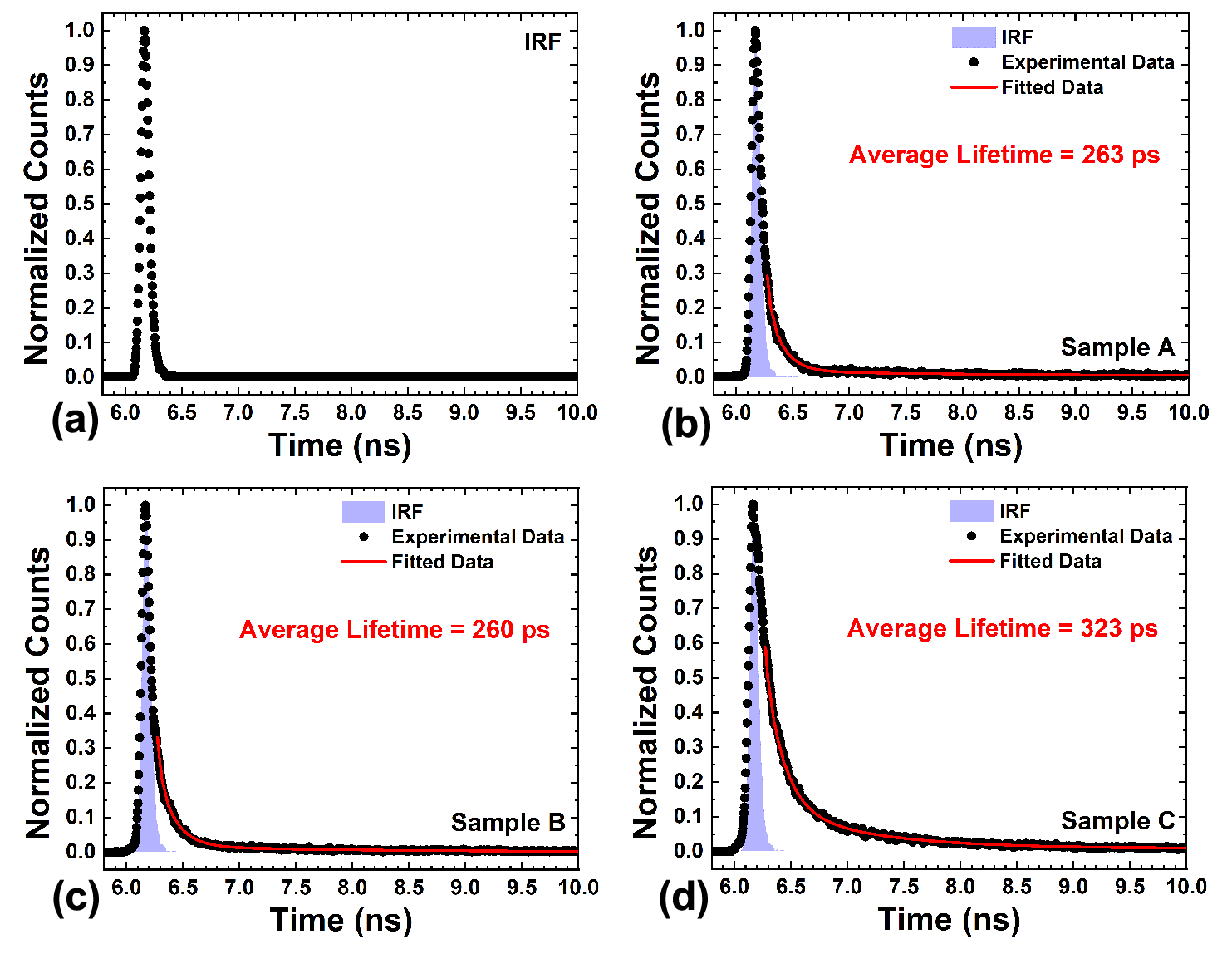}
  \caption{(a) Instrument Response Function of the TCSPC setup with an FWHM of 86 ps. TCSPC fluorescence decay transient of (b) sample A, (c) sample B, and (d) sample C}
  \label{fgr:6}
\end{figure}  

To explore the impact of alloy fluctuations on carrier dynamics, a time-resolved photoluminescence measurement was carried out in a TCSPC setup. The instrument response function (IRF) of the setup and the experimentally obtained decay profiles of the samples are shown in Figure \ref{fgr:6}. IRF-adjusted average carrier lifetimes are 263 ps, 260 ps, and 323 ps for samples A, B, and C, respectively. These values are well above the full-width half maxima (FWHM) of IRF (86 ps). 
Interestingly, sample C also exhibits an increased luminescence decay time. This further confirms the enhanced optical performance in sample C, with recent studies correlating a reduction in the average recombination time of the carriers to the presence of non-radiative centers \cite{Sheen2022, D1NR06088K}. This happens as the dislocation defects enhance the co-localization of electron-hole pairs in their vicinity through strain relaxation, unlike the individual localization of carriers near compositional fluctuations. Therefore, the higher carrier lifetime in sample C can be attributed to the superior quality of the grown structures, as already evident from the results shown in Figure \ref{fgr:4} and \ref{fgr:5}. Sample A and B, potentially having higher defect densities, exhibit faster lifetimes.

\section{Conclusion}
In conclusion, we observed that local fluctuation of Indium concentration significantly impacts the thermal redistribution of carriers in InGaN/GaN multi-quantum-disk samples. However, the typical 'S' shape temperature dependence of peak position can be less prominent in samples grown under certain conditions. Further investigation showed that a higher level of alloy fluctuation in the grown samples could also correlate with a higher density of defects impacting the luminescence intensity and the carrier dynamics.


%

\appendices


\section*{Funding}
This work is supported by the Scheme for Transformational and Advanced Research in Sciences (STARS) project (project number: STARS1/421) under the Ministry of Education, Government of India.

\section*{Acknowledgment}
We acknowledge the National Centre for Photovoltaic Research and Education at IIT Bombay for FEGSEM imaging of the samples. We would also like to thank the Industrial Research and Consultancy Centre (IRCC), IIT Bombay, for TEM imaging. We are also grateful to the IIT Bombay Nanofabrication Facility (IITBNF) team for their continuous support.

\ifCLASSOPTIONcaptionsoff
  \newpage
\fi



%
\bibliography{bibfiles}
\bibliographystyle{IEEEtran}

%

\begin{IEEEbiographynophoto}{Soumyadip Chatterjee}
received the M.Tech. degree in Electronics and Telecommunication Engineering from
Indian Institute of Engineering Science and Technology, Shibpur, Howrah, India. He is
currently pursuing a Ph.D. degree with the Department
of Electrical Engineering at IIT Bombay, Mumbai,
India.
\end{IEEEbiographynophoto}

\begin{IEEEbiographynophoto}{Subhranshu Sekhar Sahu} received his Ph.D in Physics from
National Institute of Technology, Raipur, India. He is currently an institute postdoctoral fellow at the Department of Physics, IIT Bombay, India.

\end{IEEEbiographynophoto}

\begin{IEEEbiographynophoto}{Kanchan Singh Rana}
received the M.Tech. degree from
Indian Institute of Technology Mandi, Mandi, India. He is currently pursuing a Ph.D. degree with the Department of Electrical Engineering at IIT Bombay, Mumbai, India.
\end{IEEEbiographynophoto}

\begin{IEEEbiographynophoto}{Swagata Bhunia}
received his Ph.D. degree in Physics
from the Indian Institute of Technology Bombay, Mumbai, India. He is currently working as a senior research associate at IIT Bombay.
\end{IEEEbiographynophoto}

\begin{IEEEbiographynophoto}{Dipankar Saha}
received the B.E. degree from Jadavpur University, Kolkata, India, in 2001, the M.Tech. degree from the Indian Institute of Technology Bombay, Mumbai, India, in 2005, and the Ph.D. degree from the University of Michigan, Ann Arbor, MI, USA, in 2009. He is currently a Professor at the Department of
Electrical Engineering, IIT Bombay. His current research interests include GaN-based
electronic and optoelectronic devices.
\end{IEEEbiographynophoto}


\begin{IEEEbiographynophoto}{Apurba Laha}
received the Ph.D. degree in 2004
from Indian Institute of Science, Bengaluru, India.
He was with the Institute of Electronic Material
and Device, Leibniz University Hannover, Hanover,
Germany, from 2005 to 2011. Since 2012, he has
been with IIT Bombay, Mumbai, India. He is currently a Professor of Electrical Engineering at IIT Bombay. His current research interest includes III-Nitride optoelctronics, epitaxial oxide, single photon sources, and molecular beam epitaxy.
\end{IEEEbiographynophoto}




\end{document}